\documentstyle[sprocl,11pt]{article}

\def\simge{\mathrel{%
   \rlap{\raise 0.511ex \hbox{$>$}}{\lower 0.511ex \hbox{$\sim$}}}}
\def\simle{\mathrel{
   \rlap{\raise 0.511ex \hbox{$<$}}{\lower 0.511ex \hbox{$\sim$}}}}

\def\Journal#1#2#3#4{{#1} {\bf #2}, #3 (#4)}

\def\be{\begin{equation}}
\def\ee{\end{equation}}
\def\bea{\begin{eqnarray}}
\def\eea{\end{eqnarray}}
\def\ba{\begin{array}} 
\def\ea{\end{array}}   

\begin{document}

\title{\Large \bf ABOUT ~SUPERPARTNERS  \\[.55truecm]
AND ~THE ~ORIGINS ~OF\bigskip \\
THE ~SUPERSYMMETRIC ~STANDARD ~MODEL}

\author{\vskip .2truecm \large P. FAYET}

\address{\vskip .2truecm
Laboratoire de Physique Th\'eorique de l'Ecole Normale 
Sup\'erieure\,\footnote{
UMR 8549, Unit\'e Mixte du CNRS 
et de l'Ecole Normale Sup\'erieure.
\medskip \\
}, 
\\ 
24 rue Lhomond,
75231 Paris Cedex 05, France,
\\E-mail: fayet@physique.ens.fr}

\maketitle\abstracts{\vskip .45truecm
We recall the obstacles which seemed, long ago, to prevent
supersymmetry from possibly being a fundamental symmetry of Nature.
Which bosons and fermions could be related\,?
Is spontaneous supersymmetry breaking possible\,?
Where is the spin-$\frac{1}{2}\,$ Goldstone fermion of supersymmetry\,?
Can one define conserved baryon and lepton numbers in such theories, 
although they systematically involve self-conjugate Majorana fermions\,? etc..
We then recall how an early attempt to relate the photon with a ``neutrino''
led to the definition of $\,R$-invariance, but that this ``neutrino''
had to be reinterpreted as a new particle, the {\it \,photino\,}.
This led us to the Supersymmetric Standard Model, 
involving the $\,SU(3)\times SU(2)\times U(1)\,$ gauge interactions of
chiral quark and lepton superfields, and of two doublet Higgs superfields
responsible for the electroweak breaking 
and the generation of quark and lepton masses.
The original continuous $R$-invariance 
was then abandoned in favor of its discrete version, $\,R$-parity \,-- 
reexpressed as $\ (-1)^{2 S}\ (-1)^{(3B+L)}\ \,$--\,
so that the gravitino and gluinos can acquire masses.
We also comment about supersymmetry breaking.
}


\section{Introduction}
\label{sec:intro}

\vskip .1truecm

The algebraic structure of supersymmetry in four dimensions
was introduced in the beginning of the seventies by 
Gol'fand and Likhtman~\cite{gl},
Volkov and Akulov~\cite{va}, and Wess and Zumino~\cite{wz}, as
recalled in various contributions to this book.
It involves a spin-$\frac{1}{2}\,$ fermionic symmetry generator, 
called the supersymmetry generator, 
satisfying anticommutation relations.
This supersymmetry generator $\,Q\,$ 
is defined so as to relate fermionic with bosonic fields, 
in supersymmetric relativistic quantum field theories.

\vskip .3truecm
At that time it was not at all clear if \,-- and even less how --\,
supersymmetry could actually be used to relate fermions and bosons, 
in a physical theory of particles. While very interesting 
from the point of view of relativistic field theory, 
supersymmetry seemed clearly inappropriate 
for a description of our physical world.
In particular one could not identify physical bosons and fermions 
that might be related under such a symmetry. It even seemed initially
that supersymmetry could not be spontaneously broken at all \,-- 
in contrast with ordinary symmetries --\,
which would imply that bosons and fermions be systematically degenerated 
in mass\,!
Supersymmetric theories also involve, systematically,  self-conjugate 
Majorana spinors \,-- unobserved in Nature --\, 
while the fermions that we know all appear 
as Dirac fermions carrying conserved \,($\,B\,$ and $\,L\,$)
\,quantum numbers.
In addition, how could we account for the conservation of the fermionic numbers
$\,B\,$ and $\,L\,$ 
(only carried by fermions), \,in a supersymmetric theory,
in which fermions are related to bosons\,?
Most physicists were then considering supersymmetry as irrelevant 
for ``real physics''.

\vskip .3truecm
Still this algebraic structure could actually be taken seriously 
as a possible symmetry of the physics 
of fundamental particles and interactions, once we understood 
that the above obstacles preventing the application of supersymmetry 
to the real world could be overcome. 
After an initial attempt illustrating how far one could go in trying
to relate known particles together, in particular the photon with a ``neutrino'', 
and the $\,W^\pm$ bosons with charged ``leptons'' \,--  
and the limitations of this approach --\, 
in a spontaneouly broken $\,SU(2) \times U(1)\,$ 
electroweak theory involving 
two chiral doublet Higgs superfields~\cite{R}, 
~we were quickly led to reinterpret the fermions of this
model, which all possess a conserved $\,R\,$ quantum number 
carried by the supersymmetry generator,  
\,as belonging to a new class of particles.
The ``neutrino'' ought to be considered as a really new particle,
a ``photonic neutrino'', a name which I transformed in 1977 
into {\it \,photino}, also calling at the same time {\it \,gluinos\,}
the fermionic partners of the colored gluons 
(quite distinct from the quarks!), and so on.
More generally this led us to postulate the existence of new 
$\,R$-odd ``superpartners'' 
for all ordinary particles and consider them seriously, 
despite their rather non-conventional properties:
e.g. new bosons carrying ``fermion'' number, now known 
as {\it \,sleptons} and {\it \,squarks}, or Majorana fermions 
transforming 
as an $\,SU(3)\,$ color octet, 
which are precisely the {\it \,gluinos}, etc.~\cite{ssm,grav}.
In addition the electroweak breaking must be induced by \hbox{\it a pair\,} 
of electroweak Higgs doublets, not just a single one as in the Standard Model, 
which requires the existence of {\it \,charged Higgs bosons},
\,and of several neutral ones.

\vskip .3truecm

The still-hypothetical superpartners may be 
distinguished by a new quantum number called $\,R\,$-parity~\cite{rp}, 
associated with a $\,Z_2$ remnant of the continuous $\,R$-symmetry, 
which may be multiplicatively conserved in a natural way,
and is especially useful to guarantee the absence of 
unwanted interactions mediated by squark or slepton exchanges.
The conservation (or non-conservation)
of $\,R$-parity is closely related with the conservation 
(or non-conservation) of baryon and lepton numbers, 
$\,B\,$ and $\,L\,$, \,as illustrated by the well-known formula 
reexpressing $\,R$-parity as $\,(-1)\,^{2S} \ (-1)\,^{3B+L}$
~\cite{ff}.
\,The finding of the basic building blocks of what 
we now call the Supersymmetric Standard Model 
(whether ``minimal'' or ``non-minimal'') allowed for the 
experimental searches for ``supersymmetric particles'', 
which started with the first searches for gluinos and photinos, 
selectrons and smuons, in the years 1978-1980,
and have been going on continuously since.
These searches often rely on the ``missing energy'' signature,
corresponding to energy-momentum carried away by unobserved 
neutralinos~\cite{ssm,ff,ff2,ff3}.
A conserved $\,R$-parity also ensures the stability 
of the ``lightest supersymmetric particle'',
\,a good candidate to constitute the non-baryonic Dark Matter 
that seems to be present in our Universe.
The general opinion of the scientific community towards supersymmetry and 
supersymmetric extensions of the Standard Model 
has considerably changed since the early days, 
and it is now widely admitted that supersymmetry may 
well be the next fundamental symmetry to be discovered in the physics of 
fundamental particles and interactions, 
although this remains to be experimentally 
proven.

\

\section{Nature does not seem to be supersymmetric\,!}
\label{sec:na}

Let us now travel back in time, and think 
about the supersymmetry algebra, and the way it might be realized in Nature.
This supersymmetry algebra 
\bea
\label{alg}
\cases{ \ \ \ba{cccc}
\{ \ \,Q , \ {\bar Q} \ \,\} \ \ &=& \
- \ 2\ \,\gamma_{\mu} \  P^{\mu} &, \vspace {0.3 true cm} \cr 
[ \,\ Q, \ P^{\mu} \ ] \ \ &=& 
\  \ \ \ \ \ \ \ \ \ \ \ \ 0  \ \ \ \ \ &.
\ea
}\label{ss}           
\eea
was introduced, 
in the years 1971-1973, by three different groups, 
with quite different motivations. 
Gol'fand and Likhtman~\cite{gl}, in their remarkable work published in 1971,
first introduced it 
with the apparent hope of understanding parity-violation:
when the Majorana supersymmetry generator $\,Q_\alpha\,$ 
is written as a two-component 
chiral Dirac spinor (say $\,Q_L\,$), one may have the impression that the 
supersymmetry algebra, which then involves a chiral projector 
in the right-handside of the anticommutation relation (\ref{alg}), 
is intrinsically parity-violating 
\,(which, however, is not the case); they suggested that such (supersymmetric) 
models must therefore necessarily violate parity, probably thinking
that this could lead to an explanation 
for parity-violation in weak interactions.
Volkov and Akulov~\cite{va} hoped to explain the masslessness of the neutrino 
from a possible interpretation as a spin-$\frac{1}{2}\,$
Goldstone particle, while Wess 
and Zumino~\cite{wz} wrote the algebra by extending to four dimensions the 
``supergauge'' (i.e. supersymmetry) transformations~\cite{2d}, 
and algebra~\cite{2dalg},
acting on the two-dimensional string worldsheet.
However, the mathematical existence of an algebraic structure does 
not imply that it has to play a r\^ole as an invariance 
of the fundamental laws of Nature\,\footnote{Incidentally 
while supersymmetry is commonly referred to as
``relating fermions with bosons'', \,its algebra (\ref{alg}) does not even 
require the existence of fundamental bosons\,!
(With non-linear 
realizations of supersymmetry a fermionic field can be 
transformed into a {\it \,composite\,} bosonic field made of fermionic 
ones~\cite{va}; but we shall work within the framework 
of the linear realizations of the supersymmetry algebra, 
which allows for renormalizable supersymmetric field theories.)
The supersymmetry algebra (\ref{alg}) certainly does not imply 
by itself the existence of the superpartners\,! (Just as the mathematical 
existence of the $\,SU(2)\,$ group does not imply the physical existence 
of the isospin or electroweak symmetries, 
the existence of $\,SU(3)\,$ does not imply that of the strange quark, and the
flavor or color symmetries; the existence of $\,SU(4)\,$ does not require 
technicolor, nor that of $\,SU(5)$, \,grand unification\,!)
}.

\vskip .5truecm

Indeed many obstacles seemed, long ago, to prevent supersymmetry 
from possibly being a fundamental symmetry of Nature.
Which bosons and fermions could be related by supersymmetry\,?
May be supersymmetry could act at the level of composite objects, e.g. 
as relating baryons with mesons\,?
Or should it act at a fundamental level, i.e. 
at the level of quarks and gluons\,? (But quarks are color triplets, 
and electrically charged,
while gluons transform as an $\,SU(3)\,$ color octet, 
and are electrically neutral\,!) 
\,Is spontaneous supersymmetry breaking possible at all\,?
If yes, 
where is the spin-$\frac{1}{2}\,$ Goldstone fermion of supersymmetry, 
if it cannot be identified as one of the known neutrinos\,?
Can we use supersymmetry to relate directly known bosons and fermions\,?
And, if not, why\,?
\,If known bosons and fermions cannot be directly related by supersymmetry, 
do we have to accept this as the sign that supersymmetry is {\it \,not\,}
a symmetry of the fundamental laws of Nature\,?
If we still insist to work within the framework of supersymmetry,
how could it be possible to define conserved baryon and lepton numbers 
in such theories,
which systematically involve {\it \,self-conjugate\,}
Majorana fermions, unknown in Nature, while $\,B\,$ and $\,L\,$ 
are carried only by fundamental (Dirac) fermions \,-- not by bosons~? 
And, once we are finally led to postulate the existence of new bosons 
carrying $\,B\,$ and $\,L\,$ 
\,-- the new spin-0 squarks and sleptons --\,
can we prevent them from mediating new unwanted interactions\,? 

\vskip .3truecm

While bosons and fermions should have equal masses 
in a supersymmetric theory, this is certainly not the case in Nature.
Supersymmetry should then clearly be broken.
But spontaneous supersymmetry breaking is notoriously difficult to achieve, 
to the point that it was even initially thought to be impossible\,!
Why is it so\,? 
Supersymmetry is a special symmetry, 
since the Hamiltonian, which appears in the right-handside 
of the anticommutation relations  (\ref{alg}), can be expressed 
proportionally to the sum of the squares of the components 
of the supersymmetry generator, as
$\,H\,=\,\frac{1}{4}\ \sum_\alpha\,Q_\alpha^{\ 2}\,$.
~This implies that a supersymmetry preserving vacuum state must 
have vanishing energy~\cite{iz}, while any candidate for a ``vacuum state'' 
which would not be invariant under supersymmetry
may na\"{\i}vely be expected 
to have a larger, positive, energy\,\footnote{Such a would-be supersymmetry 
breaking state corresponds, in global supersymmetry,
to a {\it \,strictly positive\,} energy density 
\,-- the scalar potential being expressed proportionally 
to the sum of the squares of the auxiliary 
$\,D, \ F\,$ and $\,G\,$ components, as
$\ \,V\,=\,\frac{1}{2}\ \sum\ (\,D^2\,+\,F^2\,+\,G^2\,)\ \,$.}. 
As a result, potential candidates for
supersymmetry breaking vacuum states
seemed to be necessarily unstable, leading to the question:
\be
\hbox{Q}1:\ \ \ \ \ 
\hbox {\it{Is spontaneous supersymmetry breaking possible at all\,?}}
\ee
As it turned out, and despite the above argument, 
several ways of breaking spontaneously 
global or local supersymmetry
have been found~\cite{fi,F,crem}.
\,But spontaneous supersymmetry breaking remains, in general,
rather difficult to obtain, since theories tend to prefer, 
for energy reasons, supersymmetric vacuum states. 
Only in very exceptional situations can the existence of 
such states be completely avoided\,!

\vskip .3truecm
As explained above in global supersymmetry a non-supersymme\-tric state has, 
in principle, 
always more energy than a supersymmetric one; 
it then seems that it should always be unstable,
the stable vacuum state being, necessarily, a supersymmetric one\,!
\,Still it is possible to escape this general result 
\,-- and this is the key to spontaneous supersymmetry breaking --\,
if one can arrange to be in one of those rare situations 
for which {\it \,no supersymmetric 
state\,} {\it exists at all\,}
\,-- the set of equations for the auxiliary field v.e.v.'s 
$\ <\!D\!>'\hbox{s}\,=\,\,<\!F\!>'\hbox{s}\,=\,<\!G\!>'\hbox{s}\,=\,0\ $ 
having {\it \,no solution at all\,}.
\,But these situations are in general 
quite exceptional. (This is in sharp contrast 
with ordinary symmetries, in particular gauge symmetries,
for which one only
has to arrange for non-symmetric states to have less energy 
than symmetric ones, in order to get spontaneous symmetry breaking.)
These rare situations usually involve an abelian $\,U(1)\,$ 
gauge group~\cite{fi}, 
allowing for a gauge-invariant linear \,``$\,\xi\,D\,$''\, 
term to be included in the Lagrangian density\,\footnote{Even 
in the presence of such a term, one frequently does not get a
spontaneous breaking of the supersymmetry: one has to be very careful 
so as to avoid the presence
of supersymmetry restoring vacuum states, which generally tend to exist.
};
\,and/or an appropriate set of chiral superfields 
with special superpotential interactions
which must be very carefully chosen (so as to get ``$F$-breaking'')~\cite{F}, 
preferentially with the help of additional symmetries such as $\,R$-symmetries.
\,In local supersymmetry~\cite{sugra}, which includes gravity,
one also has to arrange, at the price of a very severe fine-tuning,
for the energy density of the vacuum to vanish exactly~\cite{crem}, 
or almost exactly, to an extremely good accuracy,
so as not to generate an unacceptably large value of the 
cosmological constant $\,\Lambda\,$.

\vskip .5truecm

Whatever the mechanism of supersymmetry breaking, we have to get
\,-- if this is indeed possible --\,
a physical world which looks like ours (which will precisely lead
to postulate the existence of superpartners 
for all ordinary particles).
\,Of course just accepting the possibility of explicit supersymmetry breaking 
without worrying too much about the origin of 
supersymmetry breaking terms,
as is frequently done now, makes things much easier
\,-- but also at the price of introducing a large number 
of arbitrary parameters, 
coefficients of these supersymmetry breaking terms.
In any case such terms must have their origin 
in a spontaneous supersymmetry breaking mechanism, 
if we want supersymmetry to play a fundamental role, 
especially if it is to be realized as a local fermionic gauge symmetry, 
as in the framework of supergravity theories. 
We shall come back to this question of supersymmetry breaking later. 
In between, we note that the spontaneous breaking of the global supersymmetry 
must in any case generate a massless spin-$\frac{1}{2}\,$ Goldstone particle, 
leading to the next question,
\be
\hbox{Q}2: \ \ 
\hbox {\it{Where is the spin-$\frac{1}{2}\,$ 
Goldstone fermion of supersymmetry\,?}}
\ee

\noindent
Could it be one of the known neutrinos~\cite{va}\,?
A first attempt at implementing this idea 
within a $\,SU(2) \times U(1)\,$ electroweak model of ``leptons''~\cite{R}
quickly illustrated that it could not be pursued very far. 
(Actually, the ``leptons'' of this first electroweak model 
were soon to reinterpreted to become the ``charginos'' and ``neutralinos'' 
of the Supersymmetric Standard Model.)

\vskip .3truecm

If the Goldstone fermion of supersymmetry is not one of the known
neutrinos, why hasn't it been observed\,?
Today we tend not to think at all about the question,
since: 1) the generalized use of soft terms breaking {\it \,explicitly\,}
the supersymmetry seems to make this question irrelevant; \ 
2) 
since supersymmetry has to be realized locally anyway, 
within the framework of supergravity~\cite{sugra}, 
the massless \hbox{spin-$\frac{1}{2}\,$} Goldstone fermion (``goldstino'') 
should in any case be eliminated 
in favor of extra degrees of freedom for a massive 
spin-$\frac{3}{2}\,$ gravitino~\cite{grav,crem}.

\vskip .3truecm
But where is the gravitino, and why has no one ever seen 
a fundamental spin-$\frac{3}{2}\,$\, particle\,?
Should this already be taken as an argument against supersymmetry 
and supergravity theories\,? Or should one consider that the crucial test 
of such theories should be the discovery of a spin-$\frac{3}{2}\,$ particle\,?
In that case, how could it manifest its presence\,?
In fact to discuss this question properly we need to know how this 
spin-$\frac{3}{2}\,$ particle should couple to the other particles, 
which requires us to know which bosons and fermions 
could be associated under supersymmetry~\cite{ssm}.
In any case, even without knowing that, we might already anticipate 
that the interactions of the gravitino, 
being proportional to the square root of the Newton constant  
$\,\sqrt{G_N} \simeq 10^{-19}\ \,\hbox{GeV}^{-1}$, 
~should be absolutely negligible in particle physics
experiments.
Quite surprisingly this may, however, not necessarily be true\,!
We might be in a situation for which the gravitino is light, 
maybe even extremely light, so that this spin-$\frac{3}{2}\,$ particle
would still interact very much like the massless spin-$\frac{1}{2}\,$ 
Goldstone fermion 
of global supersymmetry, according to the ``equi\-valence theorem'' 
of supersymmetry~\cite{grav}.
In that case we are led back to our initial question, 
where is the spin-$\frac{1}{2}\,$ Goldstone fermion of supersymmetry\,?
But at this point we are in a position to answer, 
the direct detectability of the gravitino depending crucially on the value 
of its mass $\,m_{3/2}\,$, \,itself fixed by that of the 
supersymmetry breaking scale $\,\sqrt d \,=\, \Lambda_{ss}\,$~\cite{grav,grav2}.

\vskip .5truecm

In any case, much before getting to the Supersymmetric Standard Model,
and irrespective of the question of supersymmetry breaking, 
the crucial question,
if supersymmetry is to be relevant in particle physics, is:
\be
\hbox{Q}3: \ \ 
\hbox {\it{Which bosons and fermions 
could be related by supersymmetry\,?}}
\ee
But there seems to be no answer since known bosons
and fermions do not appear to have much in common 
\,-- excepted, maybe, for the photon and the neutrino. 
This track deserved to be explored~\cite{R},  
but one cannot really go very far in this direction.
In a more general way the number of (known) degrees of freedom 
is significantly larger for the fermions 
(now 90, for three families of quarks and leptons)
than for the bosons (27 \,for the gluons, the photon 
and the $\,W^\pm$ and $\,Z\,$
gauge bosons, ignoring for the moment
the spin-2 graviton, and the still-undiscovered Higgs boson).
And these fermions and bosons have very different gauge symmetry 
properties\,!

\vskip .3truecm

Furthermore supersymmetric theories also involve, systematically, 
self-conjugate Majorana spinors \,-- unobserved in Nature --\, 
while the fermions that we know all appear 
as Dirac fermions carrying conserved $\,B\,$ and $\,L\,$
quantum numbers.
This leads to the question
\be
\hbox{Q}4: \ \ 
\ba{c} \hbox {\it How could one define (conserved)} \\
\hbox{\it baryon and lepton numbers, 
in a supersymmetric theory ?}
\ea
\ee
These quantum numbers, presently known to be
carried by fundamental fermions only, not by bosons, 
seem to appear in Nature as {\it intrinsically-fermionic\,} numbers.
Such a feature cannot be maintained in a supersymmetric 
theory, and one has to accept the (then rather heretic)
idea of attributing 
baryon and lepton numbers to fundamental bosons, as well as to fermions.
These new bosons carrying $\,B\,$ or $\,L\,$ are
the superpartners of the spin-$\frac{1}{2}$ quarks and leptons, namely
the now-familiar (although still unobserved)
spin-0  {\it \,squarks\,} and {\it \,sleptons\,}.
Altogether, all known particles should be associated 
with new {\it \,superpartners\,}~\cite{ssm}.

\vskip .3truecm

Of course nowadays we are so used to deal with spin-0 squarks and sleptons, 
carrying baryon and lepton numbers almost by definition, 
that we can hardly imagine this could once have appeared as a problem.
Its solution went through the acceptance of the idea 
of attributing baryon or lepton numbers to a large number of new 
fundamental bosons.
But if such new spin-0 squarks and sleptons are introduced,
their direct (Yukawa) exchanges between ordinary 
 quarks and leptons, if allowed, 
could lead to an immediate disaster, preventing us from getting a theory 
of electroweak and strong interactions mediated by spin-1 
gauge bosons, and not spin-0 particles, 
with conserved $\,B\,$ and $\,L\,$ quantum numbers\,! 
This may be expressed by the following question
\be
\hbox{Q}5: \ \
\ba{c} \hbox {\it{How can we avoid unwanted interactions}} \\ 
\hbox{\it{mediated by spin-0 squark and slepton exchanges\,?}}
\ea
\ee

\noindent
Fortunately, we can naturally avoid 
such unwanted interactions, thanks to $\,R$-parity
(a remnant of the continuous $\,U(1)\ R$-symmetry) which, if present,
guarantees that squarks and sleptons can{\it not\,} be
directly exchanged between ordinary quarks and leptons, 
allowing for conserved baryon and lepton numbers 
in supersymmetric theories.

\bigskip

\section{Continuous $R$-invariance and electroweak symmetry breaking
(from an early attempt to relate the photon and the neutrino).}
\label{sec:toy}

\vskip .1truecm

Let us now return to an early attempt at relating 
{\it \,existing\,} bosons and fermions together~\cite{R}\footnote{This 
model is reminiscent of
a presupersymmetry two-Higgs-doublet model~\cite{P} 
which turned out to be very 
similar to supersymmetric gauge theories,
with Yukawa and $\,\varphi^4\,$ interactions restricted 
by a continuous $\,Q$-invariance, ancestor of the continuous $\,R$-invariance 
of supersymmetric theories.},
also at the origin of the definition of the continuous $\,R\,$-invariance
(the discrete version of which leading to $\,R$-parity).
It also showed how one can break spontaneously the $\,SU(2) \times U(1)\,$ 
electroweak gauge symmetry in a supersymmetric theory, 
using (in a modern language) a pair of chiral doublet Higgs superfields 
that would now be called $\,H_1\,$ and $\,H_2\,$. This involves 
a mixing angle initially called $\,\delta\,$ but now known as $\,\beta\,$,
defined by
\be
\tan \,\beta\ \ =\ \ \frac{v_2}{v_1}\ \ .
\ee
The fermions of this early supersymmetric model, 
which are in fact gaugino-higgsino mixtures,
should no longer be considered as lepton candidates, 
but became essentially the ``charginos'' and ``neutralinos'' 
of the Supersymmetric Standard Model~\cite{ssm,grav}.

\vskip .5truecm

Despite the general lack of similarities between known bosons and fermions, 
we tried as an exercise to see how far one could go in attempting 
to relate the spin-1 photon with a spin-$\frac{1}{2}$ neutrino.
If we want to attempt to identify the companion of the photon
as being a ``neutrino'',
despite the fact that it initially appears as a self-conjugate 
Majorana fermion, we need to understand 
how this particle could carry a conserved quantum number 
that we might interpret as a ``lepton'' number.
This was made possible through to the definition of
{\underline{\it a continuous $\,U(1)\,$ $\,R$-invariance}}~\cite{R},
which also guaranteed the masslessness of this ``neutrino'' 
(``$\nu_L$'', ~carrying $\,+1\,$ unit of $\,R\,$),
\,by acting chirally on the Grassmann coordinate $\,\theta\,$
which appears in the expression of the various 
gauge and chiral superfields.
The supersymmetry generator $\,Q_\alpha\,$ carries one unit of the 
corresponding additive conserved 
quantum number, called $\,R\,$
(so that one has $\ \Delta \,R\,=\,\pm\,1\,$  between a boson 
and a fermion related by supersymmetry).

\vskip .3truecm
Attempting to relate the photon with one of the neutrinos
could only be an exercise of limited validity.
The would-be ``neutrino'', in particular, 
while having in this model a $\,V\!-\!A\,$ coupling to its associated ``lepton''
and the charged $\,W^\pm$ boson,
was in fact what we would now call a ``photino'', 
not directly coupled to the $\,Z\,$ boson\,!
~Still this first attempt, which essentially became a part 
of the Supersymmetric Standard Model, 
illustrated how one can break spontaneously a
$\,SU(2) \times U(1)$ gauge symmetry in a supersymmetric theory, 
through an electroweak breaking induced by 
{\it a \underline{pair} \,of  chiral doublet Higgs superfields},
now known as $\,H_1\,$ and $\,H_2$ \,!
~(Using only a single doublet Higgs superfield 
would have left us with a massless charged chiral fermion, which is,
evidently, unacceptable.)
~Our previous charged ``leptons''
were in fact what we now call two winos, or charginos,
obtained from the mixing of charged gaugino and higgsino components,
as given by the mass matrix
\be
\label{mwino}
{\cal M}\ \ =\ \ \pmatrix{  (\,m_2\,=\,0\,) & 
\ \displaystyle{\frac{g\,v_2}{\sqrt 2}\,= \,m_{W}\sqrt{2}\,\sin \beta \ }\,
\cr \cr
\displaystyle{\ \frac{g\,v_1}{\sqrt 2}\,=\,m_{W}\sqrt{2}\,\cos \beta \ } 
&  \mu\,=\,0  \cr } \ \ ,
\ee

\noindent
in the absence of a direct higgsino mass that would have originated from a
$\ \mu\ H_1 H_2\ $ mass term 
in the superpotential\,\footnote{This $\,\mu\ \,H_1 H_2\,$ term, 
which would have broken explicitly 
the continuous $\,U(1)\,$ $\,R$-invariance then intended to be
associated with the ``lepton'' number conservation law, 
was already replaced by a $\ \lambda\ \,H_1 H_2\,N\ $ trilinear 
coupling involving an 
{\it \,extra neutral singlet chiral superfield\,} $\,N\,$.
}.
The whole construction showed that one could deal elegantly 
with elementary spin-0 Higgs fields
(not a very popular ingredient at the time),
in the framework of spontaneously-broken supersymmetric theories.  
Quartic Higgs couplings are no longer completely arbitrary, 
but fixed by the values of the gauge coupling constants 
\,-- here the electroweak couplings 
\,\hbox{$g\,$ and $\,g'$ --\,}\, through the following ``$D$-terms''
\,(i.e. 
\small $\ \frac{\vec D ^2}{2}\,+\,\frac{D'^2}{2}\,$
\normalsize
)\,
in the scalar potential given in ~\cite{R}
\footnote{With a different denomination for the two Higgs doublets,
such that
$\ \varphi'' \,\mapsto \,h_1,$ \ $(\varphi')^c\,\mapsto\,h_2,$ 
$\ \,\tan \delta = v'/v''\ \mapsto\ \tan \beta = v_2/v_1\ $.}:
\be
\ba{ccl}
\vspace{-.4cm}\\
\ V_{\hbox{\small{Higgs}}} \ &=&\ \displaystyle{\frac{g^2}{8}\ \ 
(\,h_1^\dagger\ \vec \tau \ h_1\,+\,h_2^\dagger\ \vec \tau \ h_2\,)^2 \ + \ 
 \frac{g'^2}{8}\ \ 
(\,h_1^\dagger\,h_1-h_2^\dagger\,h_2\,)^2 \ +\,...\ \ } \vspace{3mm}\\
&=& \ \
 \displaystyle{ \frac{g^2\,+\,g'^2}{8}\ \ 
(\,h_1^\dagger\,h_1\,-\,h_2^\dagger\,h_2\,)^2 \ \ +\ \
\frac{g^2}{2}\ \ |\,h_1^\dagger\,h_2\,|^2\ \ +\ \ ...\ \ . }
\ea
\ee

\noindent
This is precisely the quartic Higgs
potential of the ``minimal'' version of the Supersymmetric Standard Model, 
the so-called MSSM, with its quartic Higgs coupling constants equal to
\be
\frac{g^2\,+\,g'^2}{8}\ \ \ \ \hbox{and}\ \ \ \ \ \frac{g^2}{2}\ \ .
\ee
Further contributions to this quartic Higgs potential also appear
in the presence of additional superfields, such as the
neutral singlet chiral superfield $\,N\,$ already introduced in this model,
which will play an important r\^ole in the NMSSM, 
i.e. in ``next-to-minimal'' or ``non-minimal'' versions of 
the Supersymmetric Standard Model.
Charged Higgs bosons (now called $\,H^\pm$) are present 
in this framework, as well as several neutral ones.
Their mass spectrum depends on the details 
of the supersymmetry breaking mechanism considered:
soft-breaking terms, possibly ``derived from supergravity'', 
presence or absence of extra-$U(1)\,$ gauge fields 
and/or additional chiral superfields, r\^ole of radiative corrections, etc..

\bigskip

\section{The Supersymmetric Standard Model.}
\label{sec:ssm}

\vskip .1truecm
These two Higgs doublets are precisely the two doublets used in 1977 
to generate the masses of charged leptons
and down quarks, and of up quarks, in supersymmetric 
extensions of the standard model~\cite{ssm}. 
~Note that at the time 
having to introduce Higgs fields was generally 
considered as rather unpleasant.
While one Higgs doublet was taken as probably unavoidable 
to get to the standard model
or at least simulate the effects of the spontaneous breaking of the
electroweak symmetry, having to consider two Higgs doublets, 
necessitating charged Higgs bosons as well as several neutral ones,
was usually considered as a too heavy price, in addition to the 
``doubling of the number of particles'', once considered as
an indication of the irrelevance of supersymmetry.
As a matter of fact considerable work was devoted for a time on attempts 
to avoid fundamental spin-0 Higgs fields,
before returning to fundamental Higgses, 
precisely in this framework of supersymmetry.

\vskip .4truecm
In the previous $\,SU(2)\times U(1)\,$ model~\cite{R}, \,it was
impossible to view seriously for very long ``gaugino'' and ``higgsino'' fields 
as possible building blocks for our familiar lepton fields.
This led us to consider that all quarks and leptons
ought to be associated with new bosonic partners, the 
{\it \,spin-0 squarks and sleptons}.
Gauginos and higgsinos, mixed together 
by the spontaneous breaking of the electroweak symmetry,
correspond to a new class of fermions, now known as ``charginos''
and ``neutralinos''.
In particular,
the partner of the photon under supersymmetry, which 
cannot be identified with any of the known neutrinos, 
should be viewed as a new ``photonic neutrino'',
the {\it \,photino\,};
the fermionic partner of the gluon octet is an octet of self-conjugate 
Majorana fermions called {\it \,gluinos\,}, 
etc. -- although at the time {\it \,colored fermions\,} belonging to 
{\it \,octet\,} 
representations of the color $\,SU(3)\,$ gauge group were generally believed 
not to exist (to the point that one could think of using the absence of 
such particles as a general constraint to select admissible 
grand-unified theories~\cite{gm}).

\vskip .3truecm

The two doublet Higgs superfields
$\,H_1$ and $\,H_2\,$ generate 
quark and lepton masses~\cite{ssm}\,\footnote{The correspondance 
between earlier notations 
for doublet Higgs superfields, and modern ones, is as follows:
$$
 \\ [.2 true cm]
\begin{tabular}{ccc}
$S\ =\ \left( \ba{cc} S^0 \vspace{.1truecm}\\ S^-
\ea \right)\ \,\hbox{and}\ \ \,
T\ = \ \left( \ba{cc} T^0 \vspace{.1truecm}\\ T^-
\ea \right)$	 &  $\longmapsto $   &  
$H_1\ =\ \left( \ba{cc} H_1^{\,0} \vspace{.1truecm}\\ H_1^{\,-}
\ea \right)\ \,\hbox{and}\ \ \,
H_2\ = \ \left( \ba{cc} H_2^{\,+} \vspace{.1truecm}\\ H_2^{\,0}
\ea \right)\ .$
                           \\  [-.1 true cm] && \\
(left-handed) \ \ \ \ \ \ \ \ (right-handed) \  &   &   (both left-handed) 
\end{tabular}
$$
Furthermore, we originally denoted, generically, by $\,S_i\,$, left-handed, 
and $\,T_j\,$, right-handed,
the chiral superfields describing the left-handed and right-handed 
spin-$\frac{1}{2}$ quark and lepton fields, 
together with their spin-0 partners.}
\,in the usual way, through the familiar trilinear superpotential
\be
\label{supot}
{\cal W}  \ \ = \ \ h_e \ H_1 \,.\,\bar E \,L \ +\ 
h_d \ H_1\,. \,\bar D \,Q \ -\  
h_u \  H_2 \,.\,\bar U \,Q \ \ \ .			    
\ee
$L\,$ and $\,Q\,$ denote the left-handed doublet lepton and quark 
superfields, and $\,\bar E$, $\bar D\,$ and $\,\bar U\,$ left-handed singlet 
antilepton and antiquark superfields.
The vacuum expectation values of the two Higgs doublets 
described by
$\,H_1\,$ and $\,H_2\,$ generate charged-lepton and down-quark masses, 
and up-quark masses,
given by
$\,m_e\,=\,h_e\,v_1/2\,,\ \,m_d\,=\,h_d\,v_1/2\,,$ ~and 
$\,m_u\,=\,h_u\,v_2/2\,$,
~respectively.
This constitutes the basic structure of the 
{\bf \,Supersymmetric Standard Model\,},
which involves, at least, the ingredients 
shown in Table \ref{tab:basic}.
Other ingredients, such as a direct $\,\mu\ H_1 H_2\,$ direct mass term 
in the superpotential, or an extra singlet chiral superfield $\,N\,$ 
with a trilinear superpotential coupling $\ \lambda \ H_1 H_2\,N\,+ \, ... \ $
possibly acting as a replacement for a direct $\,\mu\ H_1H_2\,$ 
mass term~\cite{R},
~and/or extra $\,U(1)\,$ factors in the gauge group
(which could have been responsible for spontaneous supersymmetry breaking)
may or may not be present, depending on the
particular version of the Supersymmetric Standard Model considered.

\begin{table}[t]
\caption{\ The basic ingredients of the Supersymmetric Standard Model.
\label{tab:basic}}
\vspace{0.2cm}
\begin{center}
\begin{tabular}{|l|} \hline \\ 
\ 1) \ the  three $\,SU(3)\times SU(2)\times U(1)\,$ gauge superfield
representations;  \ \\ \\
\ 2) \ the  chiral quark and lepton superfields corresponding  \\
\hskip 2truecm to the three quark and lepton families; \\ \\
\ 3) \ the two doublet Higgs superfields $\,H_1\,$ and $\,H_2\,$
responsible  \\ 
\hskip 2truecm for the spontaneous electroweak symmetry breaking,  \\
\hskip 2truecm 
and the generation of quark and lepton masses  \\
\hskip 2truecm through 
\\  [.3truecm]
\ 4) \ \ \ \ \ \ \ \ the trilinear superpotential (\ref{supot})\ .
\\  \\ \hline
\end{tabular}
\end{center}
\end{table}

\vskip .4truecm

\vskip .3truecm

\begin{table}[t]
\caption{\ Minimal particle content of the Supersymmetric Standard Model.
\label{tab:SSM}}
\vspace{0.2cm}
\begin{center}
\begin{tabular}{|c|c|c|} \hline 
&&\\ [-0.2true cm]
Spin 1       &Spin 1/2     &Spin 0 \\ [.1 true cm]\hline 
&&\\ [-0.2true cm]
gluons ~$g$        	 &gluinos ~$\tilde{g}$        &\\
photon ~$\gamma$          &photino ~$\tilde{\gamma}$   &\\ 
------------------&$- - - - - - - - - - $&--------------------------- \\
 

$\begin{array}{c}
W^\pm\\ [.1 true cm]Z \\ 
\\ \\
\end{array}$

&$\begin{array}{c}
\hbox {winos } \ \widetilde W_{1,2}^{\,\pm} \\ 
[0 true cm]
\,\hbox {zinos } \ \ \widetilde Z_{1,2} \\ 
\\ 
\hbox {higgsino } \ \tilde h^0 
\end{array}$

&$\left. \begin{array}{c}
H^\pm\\
[0 true cm] H\ \\
\\
h, \ A
\end{array}\ \right\} 
\begin{array}{c} \hbox {Higgs}\\ \hbox {bosons} \end{array}$  \\ &&\\ 
[-.1true cm]
\hline &&

\\ [-0.2cm]
&leptons ~$l$       &sleptons  ~$\tilde l$ \\
&quarks ~$q$       &squarks   ~$\tilde q$\\ [-0.3 cm]&&
\\ \hline
\end{tabular}
\end{center}
\end{table}

In any case, independently of the details of the
supersymmetry breaking mechanism 
ultimately considered, 
we obtain the following minimal particle content
of the Supersymmetric Standard Model, given in 
\hbox{Table \ref{tab:SSM}}.
Each \hbox{spin-$\frac{1}{2}$}\, quark $\,q\,$ or charged lepton $\,l^-\,$
is associated with {\it \,two\,} spin-0 partners collectively denoted by
$\,\tilde q\,$ or $\,\tilde l^-\,$, ~while a left-handed neutrino $\,\nu_L\,$ 
is associated with a {\it \,single\,} spin-0 sneutrino $\,\tilde \nu$.
~We have ignored for simplicity 
further mixings between the various ``neutralinos''
described by neutral gaugino and higgsino fields, denoted in this table
by $\,\tilde\gamma,\ \tilde Z_{1,2}$, 
and $\tilde h^0$.
More precisely, all such models include four neutral Majorana fermions at least,
corresponding to mixings of the fermionic partners of
the two neutral $\ SU(2) \times U(1)$ gauge bosons (usually denoted by 
$\,\tilde Z\,$ and $\,\tilde\gamma$, 
~or $\,\tilde{W_3}\,$ and $\,\tilde B\,$) ~and of the 
two neutral higgsino components 
($\,\tilde{h_1^{\,0}}\,$ and $\,\tilde{h_2^{\,0}}$). 
\,Non-minimal models also involve additional 
gauginos and/or higssinos.

\

\section{About supersymmetry breaking, and the way from $\,R$-invariance 
to $\,R$-parity.}
\label{sec:grav}

\vskip .1truecm

Let us now return to the definition of the continuous 
$\,R$-symmetry, and discrete $\,R$-parity, transformations.
$\,R$-parity is associated with 
a $\,Z_2\,$ subgroup of the group of continuous 
$\,U(1)$ $\,R$-symmetry transformations, acting on the gauge superfields 
and the two doublet Higgs superfields 
$\,H_1\,$ and $\,H_2\,$ as in ~\cite{R},
with their definition extended to quark and lepton superfields
so that quarks and leptons carry $R\,=\,0\,$, 
~and squarks and sleptons, $\,R\,=\,\pm \,1\,$
(more precisely, 
$\,R=+\,1\, $ for $\,\tilde q_L,\,\tilde l_L\,$, ~and $\ R=-\,1\,$ 
for $\,\tilde q_R,\,\tilde l_R\,$)~\cite{ssm}.
As we shall see later, $R$-parity appears in fact 
as the remnant of this continuous 
$\,R$-invariance when gravitational interactions are introduced~\cite{grav},
in the framework of local supersymmetry (supergravity).
Either the continuous $\,R$-invariance, 
or simply its discrete version of $\,R$-parity, if imposed, 
naturally forbid the unwanted direct exchanges of the new 
squarks and sleptons 
between ordinary quarks and leptons.

\vskip .3truecm

These continuous $\,U(1)$ $R$-symmetry transformations, 
which act chirally on the anticommuting Grassmann coordinate
$\,\theta\,$ appearing in the definition of superspace and superfields, act
on the gauge and chiral superfields of the Supersymmetric Standard Model
as follows:

\be
\label{r}
\left\{\
\ba{cccl}
V(\,x,\,\theta,\,\bar\theta\ ) &\,\to \,&  
V (\,x, \,\theta \,e^{-i\alpha},\,\bar\theta \, e^{i\alpha}\,) &\ \, 
\hbox{for the  $\ SU(3)\times SU(2)\times U(1)\ $ gauge superfields}  
\\ [.3truecm]
H_{1,2} \,(\,x,\,\theta\,) &\to  &
H_{1,2}\, (\,x, \,\theta \,e^{-i\alpha}\,)  & \ \,
\hbox{for the left-handed doublet Higgs superfields}
 \ H_1\ \hbox{and} \ H_2  \\[.3truecm]
S (\,x,\,\theta\,) &\to  & e^{i\alpha}  \ \ 
S (\,x, \,\theta \,e^{-i\alpha}\,)  & \ \, \hbox{for the left-handed 
(anti)quark and lepton superfields} \\ [.1truecm]
&&&   \hskip 4truecm Q, \,\bar U, \,\bar D,\ \,
L, \,\bar E  \ \ .
\ea
\right.
\ee

\noindent
They are defined so as not to act on ordinary particles, 
which have $\,R=0\,$, 
\,while their superpartners have, therefore, $\,R=\,\pm1\,$.
\,They allow us to distinguish between two separate sectors
of $\,R$-even and $\,R$-odd particles. 
$\,R$-even particles 
include the gluons, photon,
$\,W^\pm\,$ and $\,Z$ gauge bosons, the various Higgs bosons, 
the quarks and leptons 
\,-- and the graviton.
$\,R$-odd particles 
include their superpartners, i.e. the gluinos 
and the various neutralinos and charginos, squarks and sleptons
\,-- and the gravitino (cf. Table \ref{tab:Rp}).
According to this first definition, $\,R$-parity simply appears as
the parity of the additive quantum number $\,R\,$,
as given by the expression~\cite{rp}:
\be
\label{rp01}
R\hbox{-parity}\ \ \,R_p\ \ =\ \ (\,-\,1\,)^{R}\ \ =\ \ \left\{ \ 
\ba{l} 
+\,1\ \ \ \ \hbox{for ordinary particles,} \vspace{2mm} \\
-\,1\ \ \ \hbox{for their superpartners.}
\ea  \right.
\ee

\vskip .5truecm
But why should we limit ourselves to the discrete $\,R$-parity
symmetry, rather than considering its full continuous parent
$\,R$-invariance \,?
This {\it \,continuous\,} $\,U(1)\,$ $\,R$-invariance, from which 
$\,R$-parity has emerged, is indeed a symmetry of all
four necessary basic building blocks 
of the Supersymmetric Standard Model~\cite{ssm}:

\vskip .2truecm
1) the Lagrangian density 
for the $\,SU(3)\times SU(2) \times U(1)\,$ gauge superfields;

\vskip .1truecm

2) the $\,SU(3)\times SU(2) \times U(1)\,$ gauge interactions 
of the quark and lepton superfields;

\vskip .1truecm

3) the $\,SU(2) \times U(1)\,$ gauge interactions 
of the two chiral doublet Higgs superfields $\,H_1\,$ and $\,H_2\,$
responsible for the electroweak symmetry breaking;

\vskip .1truecm

4) and the trilinear ``superYukawa'' interactions (\ref{supot})
responsible for quark and lepton masses.
Indeed this trilinear superpotential transforms under the continuous 
$\,R$-symmetry (\ref{r}) with ``$\,R$-weight'' $\ n_{\cal W}\,=\,\sum_i\,n_i\,=\,2\,$,
~i.e. according to
\be
{\cal W}\,(\,x,\,\theta\,) \ \ \rightarrow\ \  e^{2\,i\alpha}\ \ 
{\cal W}\,(\,x, \,\theta \,e^{-i\alpha}\,)\ \ ;
\ee

\noindent
its auxiliary ``$\,F$-component'' (obtained from the coefficient of the 
bilinear $\,\theta \,\theta \,$ term in the expansion of
$\,{\cal W}\,)$, \,is therefore $\,R$-invariant, 
generating $\,R$-inva\-riant interaction terms 
in the Lagrangian density\,\footnote{Note, however, that a direct Higgs 
superfield mass term
$\,\mu\ H_1 H_2\,$ in the superpotential, which has $\,R$-weight $\,n=0\,$,
\,does {\it \,not\,} lead to interactions which are invariant 
under the continous $\,R\,$ symmetry; but it gets in general reallowed, as 
for example in the MSSM, 
as soon as the continuous $\,R\,$ symmetry gets 
reduced to its discrete version of $\,R$-parity.}.

\begin{table}[t]
\caption{\ $R$-parities 
in the Supersymmetric Standard Model.
\label{tab:Rp}}
\vspace{0.2cm}
\begin{center}
\begin{tabular}{|c|c|} \hline 
&\\ [-0.2true cm]
Bosons       & Fermions     \\ [.2 true cm]\hline \hline 
&\\ [-0.1true cm]
$\ba{ccc}
\ba{c}
\hbox{gauge and Higgs bosons}  \\ [.1 true cm]
\hbox{graviton}
\ea
&\!&
(\,R\,=\,0\,)
\ea
$
& 
$\ba{ccc}
\ba{c}
\hbox{gauginos and higgsinos}  \\ [.1 true cm]
\hbox{gravitino}
\ea
&\!&
(\,R\,=\,\pm\,1\,)
\ea
$
\\ [-0.1 cm] & \\
$R$-parity\ \ $+$  & $R$-parity\ \ $-$ 
\\ & \\ \hline & \\[-0.1cm]
sleptons and squarks\ \ \ (\,$R\,=\,\pm\,1$\,) & 
leptons and quarks\ \ \ \ \ \ (\,$R\,=\,0$\,)  \\ [-0.1 cm] & \\
$R$-parity\ \ $-$  & $R$-parity\ \ $+$ \\ & \\ \hline
\end{tabular}
\end{center}
\end{table}

\vskip .3truecm
However, an unbroken continuous$\,R$-invariance,
which acts chirally 
on the Majorana octet of gluinos,
\be
\tilde g\ \ \to\ \ e^{\,\gamma_5\,\alpha}\ \tilde g\ \ .
\ee
would constrain them
to remain massless, even after a (spontaneous) breaking of the supersymmetry.
We would then expect the existence of relatively light
``$R$-hadrons''~\cite{ff,ff2} made of quarks, antiquarks and gluinos, 
which have not been observed. In fact we know today that gluinos, 
if they do exist, should be rather heavy,
requiring a significant 
breaking of the continuous $\,R$-invariance,
in addition to the necessary breaking of the supersymmetry.
Once the continuous $\,R$-invariance is abandoned, 
and supersymmetry is spontaneously broken,
radiative corrections do indeed allow for the generation of 
gluino masses~\cite{glu}, a point to which we shall return later.

\vskip .3truecm
Furthermore, the necessity of generating
a mass for the Majorana spin-$\frac{3}{2}\,$ {\it \,gravitino}, 
\,once {\it \,local\,} supersymmetry 
is spontaneously broken, also forces us to abandon 
the continuous $\,R$-invariance, 
in favor of the discrete $\,R$-parity symmetry, thereby 
also allowing for gluino
and other gaugino masses, at the same time as the gravitino mass $\,m_{3/2}$, 
\,as already noted in 1977~\cite{grav}.
(A third reason for abandoning the continuous $\,R$-symmetry 
could now be the non-observation at LEP
of a charged {\it wino} \,-- also called {\it \,chargino\,} --\,
lighter than the $\,W^\pm$, \,that would exist 
in the case of a continuous $\,U(1)\,$ $\,R$-invariance~\cite{R,ssm},
\,as shown by the mass matrix $\,{\cal M}\,$ of eq.\,(\ref{mwino});
the just-discovered $\,\tau^-\,$ particle could tentatively be considered, 
in 1976, as a possible light wino/chargino candidate, 
before it got clearly identified as a sequential heavy lepton.)

\vskip .5truecm
Once we drop the continuous $\,R$-invariance 
in favor of its discrete $\,R$-parity version,
we may ask how general is this notion of $\,R$-parity, and,
correlatively, are we {\it \,forced\,}
to have this $\,R$-parity conserved\,?
As a matter of fact,
there is from the beginning a close connection between $\,R$-parity 
and baryon and lepton number conservation laws, 
which has its origin in our desire 
to get supersymmetric theories in which $\,B\,$ and $\,L\,$ 
could be conserved, and, at the same time, 
to avoid unwanted exchanges of spin-0 squarks and sleptons.
Actually the superpotential of the theories discussed in Ref. \cite{ssm}
\,was constrained from the beginning, for that purpose,
to be an {\it \,even\,} function of the quark and lepton superfields.
{\it\,Odd\,} superpotential terms,
which would have violated the ``matter-parity'' symmetry 
$\,(\,-1)^{(3B+L)}$, ~were excluded,
to be able to recover $\,B\,$ and $\,L\,$ conservation laws,
and avoid direct Yukawa exchanges of spin-0 squarks and sleptons
between ordinary quarks and leptons.
Tolerating unnecessary superpotential terms which are {\it odd\ } 
functions of the quark and lepton superfields
\,(i.e. $R_p$-violating terms), 
does create, in general, immediate problems
with baryon and lepton number conservation laws
(most notably, a much too fast proton instability, if both 
$\,B\,$ and $\,L\,$ violations are simultaneously allowed).

\vskip .3truecm
This intimate connection between $\,R$-parity 
and  $\,B\,$ and $\,L\,$ conservation laws can be made quite obvious
by noting that for usual particles $\,(-1)\,^{2S}$ 
coincides with $\, \ (-1)\,^{3B+L}$, \,so that the 
$R$-parity (\ref{rp01}) may be reexpressed 
in terms of the spin $\,S\,$ and the ``matter-parity'' $(-1)\,^{3B+L}\,$, 
~as follows~\cite{ff}:
\be
\label{rp02}
R\hbox{-parity} \ \ =\ \ (-1)\,^{2S} \ (-1)\,^{3B+L}    \ \ .     
\ee

\vskip .2truecm

\noindent
This may also be written as $\ (-1)^{2S} \ (-1)\,^{3\,(B-L)}\,$, 
~showing that this discrete symmetry may still be conserved 
even if baryon and lepton numbers are separately violated,
as long as their difference ($\,B-L\,$) remains 
conserved, at least modulo 2.

\vskip .3truecm

The $\,R$-parity symmetry operator
may also be viewed as a non-trivial geometrical discrete symmetry 
associated with 
a reflection of the anticommuting fermionic Grassmann coordinate, 
$\,\theta\ \to -\,\theta\,$, in superspace~\cite{geom}.
This $\,R$-parity operator plays an essential r\^ole 
in the discussion of the experimental signatures of the new particles.
A conserved $\,R$-parity guarantees 
that {\it \,the new spin-0 squarks and sleptons 
cannot be directly exchanged\ } between ordinary quarks and leptons, 
as well as the absolute stability of the ``lightest supersymmetric particle'' 
(or LSP), a good candidate for non-baryonic Dark Matter 
in the Universe.

\vskip .5truecm 

Let us come back to the question of supersymmetry breaking. 
which still has not received a definitive answer yet.
The inclusion, in the Lagrangian density, 
of universal soft supersymmetry breaking terms 
for all squarks and sleptons,
\be
\label{soft}
-\ \sum_{\tilde q,\,\tilde l}\ m_0^{\,2}\ \ 
(\,{\tilde q}^\dagger\,\tilde q\ +\ {\tilde l}^\dagger \,\tilde l\,)\ \ ,
\ee
was already considered in 1976.
But it was also understood that such terms should in fact be 
generated by a spontaneous supersymmetry breaking mechanism, 
especially if supersymmetry is to be realized locally.
As a matter of fact they were first spontaneously generated 
with the help of the ``$D$-term'' associated with 
an {\it \,extra} $\,U(1)\,$ gauge symmetry, 
acting {\it \,axially\,} on leptons and quarks fields \,-- thereby allowing to lift 
the mass$^2$ of {\it \,both\,} ``left-handed'' and ``right-handed'' 
slepton and squark 
fields, by the same positive amount. 
When the gauge coupling constant 
$\,g"\,$ of this (still unbroken) extra $\,U(1)\,$ 
was taken to be very small, 
the supersymmetry was spontaneously broken ``at a very high scale''
$\,\sqrt d = \Lambda_{ss} \gg m_W$.   
In the limit $g" \to 0\,$, the corresponding Goldstone fermion
\,-- the gaugino of the extra $\,U(1)\,$ --\,
became completely decoupled, 
but supersymmetry was still broken with heavy slepton and squark masses; 
the breaking  was then explicit instead of spontaneous,
although only softly through the dimension 2 mass terms
\,(\ref{soft}).

\vskip .3truecm

To get a true spontaneous breaking of the supersymmetry 
with a physically coupled goldstino (of course to be subsequently 
``eaten'' by the 
spin-$\frac{3}{2}\,$ gravitino)
rather than an explicit (although soft) one, 
as well as a spontaneous breaking of the extra $\,U(1)\,$ symmetry,
and also to render, at the same time, the superpotential (\ref{supot}) 
invariant 
under this extra $\,U(1)\,$ symmetry so that it can actually be responsible 
for the generation of lepton and quark masses, 
we modified the definition 
of this extra $U(1)\,$  so that it also acts on the Higgs superfields 
$\,H_1\,$ an $\,H_2\,$
as well as on lepton and quark superfields, as follows:
\be
\label{extra}
\!\!\left\{ \ \ 
\ba{ccccl}
V(\,x,\,\theta,\,\bar\theta\ )&\to &  
V(\,x,\,\theta,\,\bar\theta\ )&&
\hbox{\ for the $SU(3) \times SU(2) \times U(1) $ gauge superfields;}  
\\ [.25truecm]
H_{1,2} (x,\,\theta) & \to  &
\ e^{-\,i\alpha}\ H_{1,2} (x,\,\theta)  &\ & 
\hbox{\ for the left-handed doublet Higgs superfields} \ \,
H_1\ \hbox{and}\ H_2  
\\  [.25truecm]
S (x,\,\theta) &\ \to \   & e^{i\frac{\alpha}{2}}  \ 
S (x,\,\theta)  && 
\hbox{\ for the left-handed (anti)quark and (anti)lepton} \\ 
&&&& \hskip 2truecm  \hbox{superfields}\ \ \ \ \ Q, \,\bar U, \,\bar D,\ \,
L, \,\bar E\ \ .
\ea
\right. 
\ee
This newly-defined extra $\,U(1)\,$ (acting on the two Higgs doublets
so that it gets spontaneously broken together with the electroweak 
symmetry), 
is now a symmetry of the trilinear superpotential interactions (\ref{supot}), 
so that lepton and quark can now acquire masses 
in a way compatible with the spontaneous supersymmetry breaking 
mechanism used.
This extra $\,U(1)\,$ is associated, 
in the simplest case of eq.\,(\ref{extra}),
with a purely axial extra $\,U(1)\,$ current 
for all quarks and charged leptons.
Gauging such an extra $\,U(1)\,$, 
which must in any case be different from the weak-hypercharge $\,U(1)\,$, 
is in fact necessary~\cite{extrau},
\,if one intends to generate large {\it \,positive\,} mass$^2$ 
for {\it \,all\,} squarks ($\,\tilde u_L,\,\tilde u_R,\ \tilde d_L,\,\tilde d_R\,$)
and sleptons, at the classical level, in a spontaneously-broken
globally supersymmetric theory (otherwise we could not avoid 
squarks having negative or at best very small mass$^2$).
But this method of spontaneous supersymmetry breaking also led to several 
difficulties. In addition to the question of anomalies, 
it required new neutral current interactions beyond those of the Standard Model.
This was fine at the time, in 1977, but such interactions
did not show up while the $\,SU(2) \times U(1)\,$ neutral current structure 
of the Standard Model got experimentally confirmed. This mechanism also 
left us with the question of generating large gluino masses.
Altogether, the gauging of an extra $U(1)\,$ no longer appears
as an appropriate way to generate large superpartner masses.
One now uses again, in general, soft supersymmetry-breaking 
terms~\cite{gg} generalizing those of eq.\,(\ref{soft}) 
\,-- possibly ``induced by supergravity'' --\,
which essentially serve as a parametrization 
of our ignorance about the 
true mechanism of supersymmetry breaking chosen by Nature
to make superpartners heavy.

\vskip .3truecm

Let us return to gluino masses. As we said before continous
$\,R$-symmetry transformations act {\it \,chirally\,} 
on gluinos, so that an unbroken $\,R$-invariance
would require them to remain massless,
even after a spontaneous breaking of the supersymmetry\,!
Thus the need, once it became experimentally 
clear that massless or even light gluinos could not be tolerated, 
to generate a gluino mass either from radiative
corrections~\cite{glu}, or from supergravity (see already {\cite{grav}), 
with, in both cases, the continuous $\,R$-invariance reduced to its 
discrete $\,R$-parity subgroup.

\vskip .3truecm
In the framework of global supersymmetry
it is not so easy to generate large gluino masses.
Even if global supersymmetry is spontaneously broken, 
and if the continuous $R$-symmetry is not present, 
it is still in general rather difficult to obtain large masses for gluinos, 
since: \ 
{\bf i)} \ no direct gluino mass term is present in the Lagrangian density; 
and \ 
{\bf ii)} \ no such term may be generated spontaneously, at the tree 
approximation, gluino couplings involving {\it colored\,} spin-0 fields. 
A gluino mass may then be generated by radiative corrections
involving a new sector of quarks sensitive 
to the source of supersymmetry breaking~\cite{glu},
that would now be called  ``messenger quarks'',
but \ {\bf iii)} \ this can only be through diagrams which ``know'' both about:
\, {\bf a)} \, the spontaneous breaking of the global supersymmetry,
through some appropriately-generated v.e.v's for auxiliary components,
$\,<\!D\!>,\ <\!F\!>\,$ or $\,<\!G\!>\,$'s;\ \ 
\, {\bf b)} \, the existence of superpotential interactions 
which do not preserve 
the continuous $\,U(1)\,$ $\,R$-symmetry.
Such radiatively-generated gluino masses, however, 
generally tend to be rather small, unless one introduces, in 
some often rather complicated ``hidden sector'', 
very large mass scales $\,\gg\,m_W\,$.

\vskip .5truecm

     Fortunately gluino masses may also result directly from 
supergravity, as already observed in 1977~\cite{grav}. Gravitational 
interactions require, within local supersymmetry, that the spin-2 
graviton be associated with a spin-$3/2\,$ partner~\cite{sugra}, the 
gravitino. Since the gravitino is the fermionic gauge particle of 
supersymmetry it must acquire a mass, $\,m_{3/2} \ (= \kappa \ d/ \sqrt{6}
 \ \approx \,
d/m_{\rm{Planck}}\,\,)$, as soon as the local supersymmetry gets spontaneously 
broken. 
Since the gravitino is a self-conjugate Majorana fermion
its mass breaks the continuous $\,R$-invariance which acts chirally on it,
just as for the gluinos, 
\,forcing us to abandon the continuous $U(1)$ $R$-invariance, 
in favor of its discrete $R$-parity subgroup.
In particular, in the presence of a spin-$\frac{3}{2}\,$ gravitino mass 
term $\,m_{3/2}\,$,
~which corresponds to $\ \Delta \,R\,=\,\pm \,2\,$,
~the ``left-handed sfermions''
$\,\tilde f_L$, ~which carry $\,R\,=\,+\,1$, ~can mix with the 
right-handed'' ones  $\,\tilde f_R$, 
~carrying $\,R\,=\,-\,1$, ~through mixing terms having 
$\ \Delta \,R\,=\,\pm \,2\,$, ~which may naturally \,(but not necessarily)
be of order $\ m_{3/2}\ m_f\,$.
Supergravity theories offer, in addition, a natural framework 
in which to include direct gaugino Majorana mass terms 
\be
-\ \frac{i}{2}\ \ m_3\ \ \bar {\tilde G}_a\,\tilde G_a\ \
-\ \frac{i}{2}\ \ m_2\ \ \bar {\tilde W}_a\,\tilde W_a\ \
-\ \frac{i}{2}\ \ m_1\ \ \bar {\tilde B}\,\tilde B\ \ ,
\ee
which also correspond to $\,\Delta \,R\,=\,\pm \,2\,$.
~The mass parameters $m_3,\ m_2\,$ and $\,m_1$, 
~for the $\,SU(3) \times SU(2) \times U(1)\,$ gauginos,
could naturally \,(but not necessarily)\,
be of the same order as the gravitino mass $\,m_{3/2}\,$.
\,Incidentally,  once the continuous $R$-invariance 
is reduced to its discrete $R$-parity subgroup, 
a direct Higgs superfield mass term $\ \,\mu \ H_1 H_2\,$,
~which was not allowed by the continuous $U(1)\,$
$\,R$-symmetry, gets reallowed in the superpotential,
as for example in the MSSM. The size of this 
supersymmetric $\,\mu\,$ parameter 
(which breaks explicitly both the continuous $\,R$-invariance (\ref{r})
and the (global) extra $\,U(1\,$) symmetry (\ref{extra}) 
may then be controlled by considering one or the other of these two symmetries.
In general, irrespective of the supersymmetry breaking mechanism
considered, one normally expects the various superpartners not to be too heavy, 
otherwise the corresponding new mass scale would tend to 
contaminate the electroweak scale, thereby
{\it creating\,} a hierarchy problem in the Supersymmetric Standard Model.
Superpartner masses are then normally expected to be naturally of the
order of $\,m_W$, ~or at most in the $\ \sim\,$ TeV$/c^2\,$ range.

\vskip .3truecm

\vskip .1truecm

The Supersymmetric Standard Model (``minimal'' or not),
with its $R$-parity symmetry (absolutely conserved, or not),
provided the basis for the experimental searches for the new superpartners
and Higgs bosons, starting with the first searches for gluinos and photinos, 
selectrons and smuons, at the end of the seventies. 
How the supersymmetry should actually be broken,
if indeed it is a symmetry of Nature, is not known yet.
Many good reasons to work on the Supersymmetric Standard Model 
and its various extensions have been discussed, 
dealing with supergravity, the high-energy unification 
of the gauge couplings, extended supersymmetry, new spacetime dimensions,
superstrings, \hbox{``$M$-theory'',} ... .
However, despite all the efforts made for more than twenty years 
to discover the new inos and sparticles, 
we are still waiting for experiments to disclose the missing 
half of the SuperWorld\,!

\end{document}